\documentclass[pre,aps,floats,superscriptaddress,floatfix,twocolumn]{revtex4}
\usepackage{amssymb,amsmath}
\usepackage{amsmath,amssymb}
\usepackage{graphicx}
\usepackage{psfrag}
\usepackage{color}
\usepackage{soul}
\usepackage{dcolumn}
\usepackage[utf8]{inputenc}
\usepackage{hyperref}

\usepackage{bm}
\usepackage[normalem]{ulem}
 \hypersetup{
    colorlinks=true,
    citecolor=blue,
    }
 \usepackage{amsmath}

\begin{document}

\title{Stationary densities in a weakly nonconserving asymmetric exclusion processes with finite resources}
\author{Sourav Pal}\email{isourav81@gmail.com}
\affiliation{Theory Division, Saha Institute of
Nuclear Physics, A CI of Homi Bhabha National Institute, 1/AF Bidhannagar, Calcutta 700064, West Bengal, India}
\author{Abhik Basu}\email{abhik.123@gmail.com, abhik.basu@saha.ac.in}
\affiliation{Theory Division, Saha Institute of
Nuclear Physics, A CI of Homi Bhabha National Institute, 1/AF Bidhannagar, Calcutta 700064, West Bengal, India}

\begin{abstract}
Asymmetric exclusion process (TASEP) along a one-dimensional (1D) open channel sets the paradigm for 1D driven models and nonequilibrium phase transitions in open 1D models. Inspired by the phenomenologies of an open TASEP with Langmuir kinetics (Lk) and with finite resources, we study the stationary densities and phase transitions in a TASEP with Lk connected to a particle reservoir at its both ends. We calculate the stationary density profiles and the phase transitions. The resulting phase diagrams in the plane of the control parameters are significantly different from their counterparts in an open TASEP with Lk. In particular, some of the phases admissible in the open TASEP with Lk model are no longer possible. Intriguingly, our model that is closely related to a TASEP coupled with Lk on a ring with a point defect, admits more phases than the latter. Phenomenological implications of our results are discussed.
\end{abstract}

\maketitle

\section{INTRODUCTION}
 \label{intro}
 
 Transport processes are integral to a wide variety of natural and engineered systems operating far from equilibrium. From molecular motors moving along cytoskeletal filaments to vehicles flowing through traffic networks and pedastrian flow, understanding how collective motion emerges under continuous energy input remains a central challenge in nonequilibrium statistical physics. Among the minimal models that capture the essential features of driven transport, the totally asymmetric simple exclusion process (TASEP) has been instrumental, see Refs.~\cite{spohn1991,driven-diff4,mukamel2000,spschutz2001} for general reviews on TASEP. Originally introduced as a simple stochastic model for biopolymerization dynamics~\cite{macdonald}, the TASEP has since emerged as a paradigmatic nonequilibrium model, owing to its boundary-induced phase transitions and the occurrence of phase coexistence even in one-dimensional (1D) systems~\cite{krug,krug1,krug2,derrida,blythe}. A TASEP consists of particles hopping unidirectionally on a one-dimensional lattice with hard-core exclusion, connected at both ends to particle reservoirs without any spatial extension or internal dynamics. Subsequently, numerous modifications and extensions of the TASEP have been developed to capture features relevant to real-life transport processes. Motivated by the dynamics of molecular motor transport along complex microtubule networks in biological cells~\cite{howard}, the TASEP has been studied on simple network geometries, where the conditions for the emergence of different phases and domain walls on various network segments have been systematically explored~\cite{andrea1,rakesh1,rakesh2}. Related studies showed that molecular motors of the kinesin-8 family can regulate microtubule length by controlling the kinetics of microtubule depolymerization~\cite{melbinger1,melbinger2}. Moreover, systems consisting of two TASEP lanes coupled by rare particle-switching events exhibit a variety of complex behaviors, including domain-wall delocalization and jam formation~\cite{reichenbach1,reichenbach2,reichenbach3}. Another key factor in many transport systems is the presence of obstacles that impede or temporarily block particle movement. For instance, vehicles on a road may be slowed or brought to a halt by other vehicles or traffic lights~\cite{helbing2001,kerner1998,treiber2013}. Similarly, in cellular transport, motor proteins such as kinesin and dynein move along microtubules to carry cargos, but their motion can be hindered by microtubule-binding proteins~\cite{dixit2008,telley2009}. These inhomogeneities along the track may be static, encompassing both localized slow sites and spatially extended bottlenecks or slow segments~\cite{lebo,niladri-tasep,kolomeisky1998,qiu2007,sm-tasep-atri3,pal-point-defect,shaw2004,mustansir,hinsch,tirtha-niladri,parna-anjan,tirtha-qkpz,pierobon2006}; or they may be dynamic~\cite{klumpp2005,waclaw2019,jindal2020,sahoo2014}, appearing and disappearing stochastically at different sites. How space dependent hopping rates can give rise to spatially nonuniform stationary densities have been studied in Refs.~\cite{tirtha-qkpz,sm-tasep-atri1,sm-ab-tasep}.

 An isolated TASEP necessarily conserves particle numbers in its bulk. In a novel generalization of TASEP, studies have been performed to explore the consequence of weak particle nonconservation in the bulk of the TASEP. The classic example is where the TASEP is coupled in its bulk with a particle reservoir by allowing particles to attach to or detach from bulk sites, a mechanism commonly known as Langmuir kinetics (Lk). The competition between the nonequilibrium directed transport and the equilibrium adsorption–desorption processes gives rise to rich behavior, including localized shocks and phase coexistence~\cite{ef-lktasep-prl,ef-lktasep-pre}. These phenomena have been quantitatively explained through phenomenological domain-wall theory~\cite{popkov2003,evans2003,levine2004,juhasz2004}, and similar jammed or shock-like structures have also been observed in biological motor transport~\cite{traffic-exp,nishinari2005,klumpp2003,vilfan2001} and vehicular traffic models~\cite{schadschneider2002,chowdhury2000,helbing2001,nagel1992}. Subsequently, a TASEP on an inhomogeneous ring coupled to Lk has also been explored~\cite{tirtha-lk1,tirtha-lk2}, shedding valuable insight on how the competition between defects, ring geometry and Lk ultimately control the spatially nonuniform stationary densities and various phase transitions. Furthermore, studies have shown that incorporating next–nearest–neighbor interactions can generate additional collective effects, including reentrance transitions and traffic-like flow patterns~\cite{antal2000}. A detailed mean-field investigation of the TASEP with Langmuir kinetics has also uncovered dynamical transitions and nontrivial relaxation behaviour arising from the competition between bulk exchange and directed motion~\cite{botto2019}. Complementarily, study of intracellular motor transport illustrate how attachment–detachment kinetics, defects, and crowding naturally give rise to jamming and shock-like structures in biological systems~\cite{appert-rolland2015}. 
 
 Other interesting generalizations of TASEP include coupling with 1D diffusion or symmetric exclusion processes in open~\cite{sm-tasep-atri2} or half closed geometries~\cite{erwin-sep-tasep1,erwin-sep-tasep2}, which show how complex space-dependent stationary density profiles emerge as consequences of the interplay between TASEP and SEP in these models. It is now understood on the basis of investigations of two-lane systems with symmetric or asymmetric lane-changing  that the steady-state behavior is strongly influenced not only by the magnitudes but also the ratios of inter-lane rates~\cite{jiang2007,gupta2014,wang2007,dhiman2014pre,dhiman2014epl}. Earlier work by Wang \textit{et al.}~\cite{wang2007-twolane,wang2008-inhomo} considered finite-size two-lane TASEPs with Lk, with and without local defects. They reported that lane-changing can induce “jumping” domain-wall shifts, with local inhomogeneities propagating across lanes, whose influence is suppressed or amplified depending on the strengths of Lk and inter-lane coupling. More recent developments have explored extended or occupancy-dependent Lk, where the attachment/detachment rates on one lane depend on the occupation of the corresponding site on the other lane. Tamizhazhagan \textit{et al.}~\cite{tamizhazhagan2023} demonstrated that this cross-lane coupling modifies phase boundaries, creates new boundary layers, and introduces finite-size effects not seen in simple Lk models. Similarly, a study in Ref.~\cite{extendedlk2022} using cluster mean-field theory confirmed that allowing Lk rates to depend on the other lane’s occupancy produces distinct correlations and novel steady states for unidirectional versus bidirectional flows.
 
 Recently, there has been a growing interest in understanding transport phenomena under resource limitations, which necessitates the study of TASEP models beyond the conventional open-boundary setup with an infinite particle supply. These studies are motivated by a variety of real-life contexts. In cellular biology, protein synthesis is limited by the finite pool of ribosomes available to translate mRNA~\cite{reser1,lim-bio1}, whereas in traffic dynamics, TASEP-like models capture vehicle movement within networks containing only a restricted number of vehicles~\cite{traffic1,limited,lim-bio2}. Applications to driven diffusive systems with constrained particle availability have also been explored~\cite{lim-driv}, and experimental validation comes from studies of spindle dynamics in eukaryotic cells~\cite{lim-exp1,lim-exp2}. A natural starting point for finite-resource modeling is a single TASEP lane coupled to a reservoir with a limited particle supply. In such systems, the entry rate becomes a function of the instantaneous reservoir population, while the exit rate remains constant. Extensions to multiple TASEP lanes competing for a common reservoir have revealed rich nonequilibrium behaviour, including phase coexistence and shock formation, as demonstrated through mean-field theory and large-scale Monte Carlo simulations~\cite{reser1,reser3,reser2}. Beyond particle-number limitation alone, a complementary line of research considers simultaneous constraints on both the hopping particles and the ``fuel carriers'' that supply the energy required for directed motion~\cite{brackley}. This dual limitation leads to multiple phase transitions as the entry and exit rates are varied. A related approach incorporates explicit dependence of \emph{both} entry and exit rates on the current reservoir population, giving rise to dynamics-induced competition between these rates and revealing \textit{reservoir crowding effects} on the steady state~\cite{astik-1tasep}, which has been subsequently generalized to models with two TASEP lanes connected antiparellely to two particle reservoirs~\cite{astik-erwin,sourav-3}. Reservoir crowding has also been explored in the presence of dynamic defects~\cite{arvind1}. In a similar vein, Refs.~\cite{sourav-1,sourav-2} investigate how inter-reservoir diffusion, together with global particle number conservation, governs the steady-state behavior of a TASEP-reservoir system under conditions of limited particle availability, as well as when this constraint is relaxed. The interplay between two reservoirs coupled to TASEPs~\cite{arvind2,arvind3} and the generalization to $K$-exclusion TASEPs~\cite{seppa}, where each site can hold up to $K$ particles, have further advanced the study of limited-resource systems.

In this article, we systematically investigate how the steady-state behaviors, e.g., the stationary density profiles and phase diagrams, are shaped by the interplay between the directed transport in the TASEP, and an equilibrium Lk adsorption--desorption process, together with  finite resources that control the effective entry and exit rates of the TASEP lane. We show that the ensuing phase diagrams in the plane of the TASEP entry and exit rate parameters have structures fundamentally different  from those in other related models of TASEP.  These differences are essentially consequences of the interplay between the weak nonconserving Lk and the constraints on TASEP due to the finite resources and the associated reservoir-controlled effective entry and exit rates in the TASEP lane.  Our work should served as a starting point for physical basis in a wide-ranging phenomena, e.g., movement of molecular motors with finite processivity along filopod extension~\cite{howard2001,kruse2002} and traffic models incorporating bulk on--off ramps~\cite{chowdhury2000}. The remainder of the manuscript is organized as follows. In Section~\ref{model}, we introduce the model and its dynamical rules, followed by a brief overview of the open boundary TASEP in Section~\ref{tasep}. The steady-state phase diagrams obtained from both mean-field theory and Monte Carlo simulations are presented in Section~\ref{pd}. In Section~\ref{mft}, we provide a detailed mean-field analysis of the steady-state densities, phase boundaries, and phase transitions, and explain the absence of certain phases in the phase diagram. Finally, we conclude in Section~\ref{summary} with a summary of our results and an outlook for future work. For completeness, the equations of motion and the particle-hole symmetry of the model are discussed in Appendix~\ref{ph-symmetry}.



 \section{THE MODEL}
 \label{model}
 
\begin{figure}[!h]
 \centerline{
 \includegraphics[width=\linewidth]{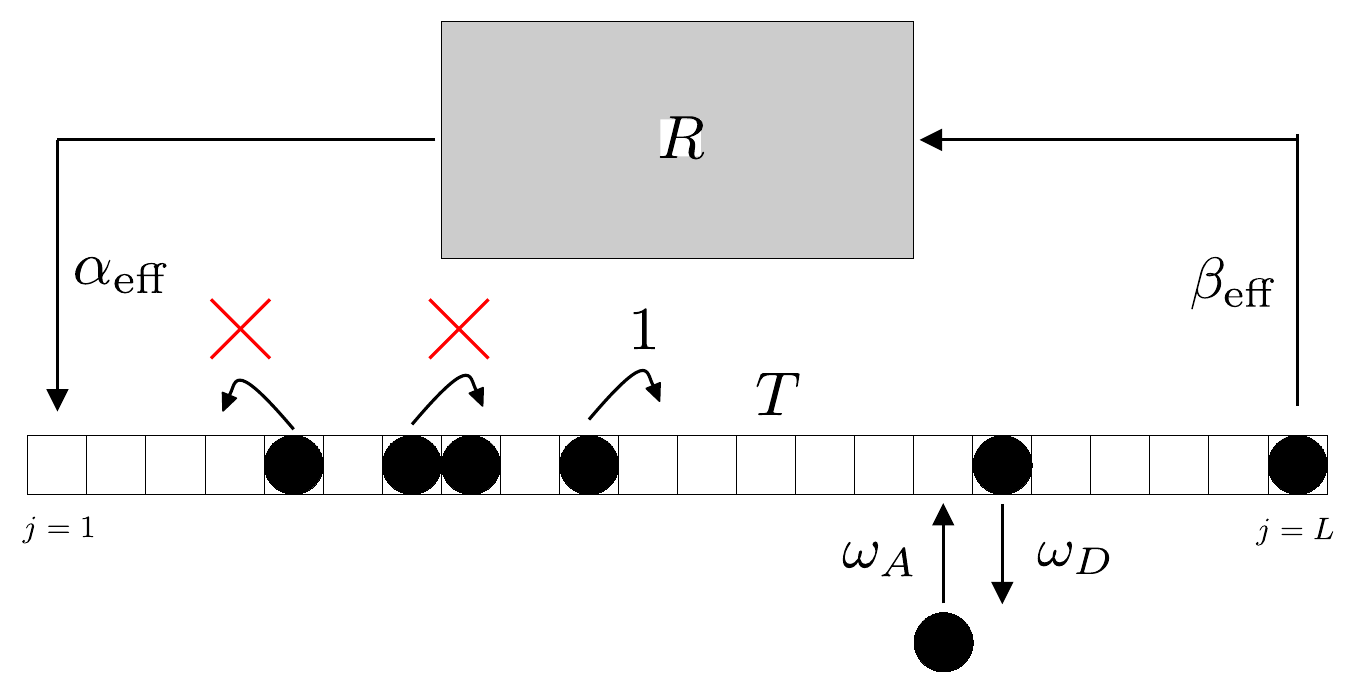}}
 \caption{\textbf{Schematic diagram of the model.} A one-dimensional lattice $T$ having $L$ sites ($j = 1, 2, \ldots, L$) executing TASEP dynamics is coupled to a particle reservoir $R$ at both boundaries ($j=1,\,L$). Each site can hold at most one particle, while the reservoir population is bounded by $L$. Particles enter at site $j=1$ with rate $\alpha_{\text{eff}}$ and exit from site $j=L$ with rate $\beta_{\text{eff}}$, as defined in Eq.~\eqref{effective-entry-and-exit-rates}, both depending on the instantaneous reservoir occupation $N_R$. Particles at sites $1 \le j < L$ hop to the next site $j+1$ if it is empty, with the hopping rate set to $1$ to fix the time scale. Additionally, within the bulk ($1 < j < L$), particles may attach to an empty site or detach from an occupied site with rates $\omega_A$ and $\omega_D$, respectively, as specified in Eq.~\eqref{attachment-detachment-rates}.
}
 \label{model-diagram}
 \end{figure}
 
We consider an exclusion process on a one-dimensional lattice $T$ with $L$ sites ($j = 1, 2, \ldots, L$), where particles are exchanged with a reservoir $R$ at both boundaries ($j=1$ and $j=L$); see Fig.~\ref{model-diagram} for a schematic diagram of the model. Each site can accommodate at most one particle at any given time; thus the occupation number $n_j$ of site $j$ can take only two values, $0$ (empty) or $1$ (occupied). Devoid of any spatial structure or internal dynamics, the reservoir functions solely as a localized source and sink, coupling to the TASEP only through particle entry and exit processes. In contrast to the open-boundary TASEP, which is coupled to reservoirs with unlimited resources, the reservoir in our model operates under a finite-resource constraint, see Refs.~\cite{reser1,reser2,reser3,pal-point-defect,limited,lim-driv,brackley,astik-1tasep,astik-erwin,arvind1,sourav-1,sourav-3}. This constraint is encoded in the effective entry and exit rates, $\alpha_{\mathrm{eff}}(N_{R})$ and $\beta_{\mathrm{eff}}(N_{R})$, which depend explicitly on the instantaneous number of particles $N_{R}$ remaining in the reservoir. We parametrize the effective entry and exit rates in terms of the bare rates $\alpha$ and $\beta$, which are both positive definite:
\begin{equation}
\label{effective-entry-and-exit-rates}
  \alpha_{\mathrm{eff}}(N_R) = \alpha\, f(N_R),
  \qquad
  \beta_{\mathrm{eff}}(N_R) = \beta\, \bigl[1 - f(N_R)\bigr].
\end{equation}
Here, the reservoir feedback function $f(N_{R})$ must be a monotonically increasing function of $N_{R}$, since a larger reservoir population should facilitate particle entry into the TASEP while simultaneously suppressing particle exit from it. Although various forms of such monotonic functions have been explored in the literature~\cite{reser1,reser2,reser3,pal-point-defect,brackley,astik-1tasep,astik-erwin,arvind1,sourav-1,sourav-2,sourav-3}, we adopt the simple form employed in Refs.~\cite{pal-point-defect,astik-erwin,sourav-1,sourav-3}, namely,
\begin{equation}
 \label{f}
 f(N_{R}) = \frac{N_{R}}{L}.
\end{equation}
Since the physical rates $\alpha_{\mathrm{eff}}(N_R)$ and $\beta_{\mathrm{eff}}(N_R)$ are non-negative, it follows that $0 \le f(N_R) \le 1$. This, in turn, imposes an upper bound on the reservoir population, i.e., $N_R \le L$. Apart from particle exchange with the reservoir through the boundary sites $j=1$ and $j=L$, particles move within the bulk of the lattice according to the usual TASEP dynamics. A particle hop from site $j$ to $j+1$ (with $1 \le j < L$) occurs at rate $1$ provided that site $j$ is occupied while site $j+1$ is empty.

 In addition to the conserved driven-transport dynamics discussed above, we also introduce Langmuir kinetics (Lk) which is no longer particle-conserving in the bulk: particles may attach to an empty site or detach from an occupied site for $1< j < L$ at equal rate $\omega$. In the thermodynamic limit ($L \to \infty$, where $L$ is the size of the TASEP), if $\omega$ is ${\cal O}(1)$, the Lk dynamics dominate. Nontrivial collective behavior arises only when the time scales of Lk and TASEP hopping compete. Following Ref.~\cite{ef-lktasep-pre,tirtha-lk1}, we therefore scale the local Lk rates as
\begin{equation}
\label{attachment-detachment-rates}
\omega = \frac{\Omega}{L}, 
\end{equation}
where $\Omega\sim\mathcal{O}(1)$. Thus the actual attachment and detachment rates scale as $1/L$, which makes them competing with the hopping process of TASEP. 
Our model has three control parameters -- $\alpha$, $\beta$ and $\Omega$. 

Our model significantly differs from an open TASEP with Lk in having the TASEP entry and exit rates being dynamically controlled by the instantaneous reservoir population, whereas for an open TASEP these are the control parameters. This results into  phase diagrams significantly different from those in the open TASEP with Lk model~\cite{ef-lktasep-prl,ef-lktasep-pre}. Furthermore,
our study also generalizes the work reported in Ref.~\cite{tirtha-lk1}, where a TASEP in a periodic system with a point defect 
coupled with Lk dynamics~\cite{tirtha-lk1}. In Ref.~\cite{tirtha-lk1} the hopping rate across the point defect is less than unity, whereas it is unity elsewhere in the ring. The presence of the point defect manifestly breaks translation invariance along the ring, allowing for macroscopically nonuniform stationary density profiles. Likewise in the present study, the particle reservoir breaks the translation invariance along the ring and, as we shall see below, gives rise to nonuniform steady states and phase diagrams more complex than those found in Ref.~\cite{tirtha-lk1}.  

 \section{TASEP}
 \label{tasep}
 
 Before we study our model, it is useful to briefly revisit an isolated open TASEP and the steady states it admits. An open TASEP consists of a one-dimensional lattice of $L$ number of sites. Particles enter the first site at rate $\alpha$ whenever it is empty, hop to the right provided the target site is vacant, and exit from the last site at rate $\beta$. The bulk hopping rate is chosen to be unity without loss of generality, since any other choice can be absorbed into a simple rescaling of time. These simple dynamical rules nevertheless give rise to rich nonequilibrium behavior. Depending on the boundary rates, the TASEP displays three steady-state phases with spatially uniform densities: an LD phase for $\alpha<\beta,\,\alpha<1/2$, an HD phase for $\beta<\alpha,\,\beta<1/2$, and an MC phase for $\alpha,\,\beta>1/2$. Each phase is characterized by distinct bulk densities and steady currents: $\rho_{\mathrm{LD}}=\alpha<1/2$ with $J_{\mathrm{LD}}=\alpha(1-\alpha)<1/4$ in the LD phase; $\rho_{\mathrm{HD}}=1-\beta>1/2$ with $J_{\mathrm{HD}}=\beta(1-\beta)<1/4$ in the HD phase; and $\rho_{\mathrm{MC}}=1/2$ with $J_{\mathrm{MC}}=1/4$ in the MC phase. In addition, along the line $\alpha=\beta<1/2$ the system develops a macroscopic spatial nonuniformity in the form of a domain wall (DW) separating the LD and HD regions. The domain wall in this DW phase is \textit{delocalized} as a consequence of the particle nonconserving boundary dynamics of the open TASEP. See Refs.~\cite{krug,krug1,krug2,derrida,blythe} for a general overview of TASEP.
 
 \section{THE PHASE DIAGRAMS}
 \label{pd}

 \begin{widetext}

 \begin{figure}[ht]
 \begin{minipage}{0.32\linewidth}
\includegraphics[width=\textwidth]{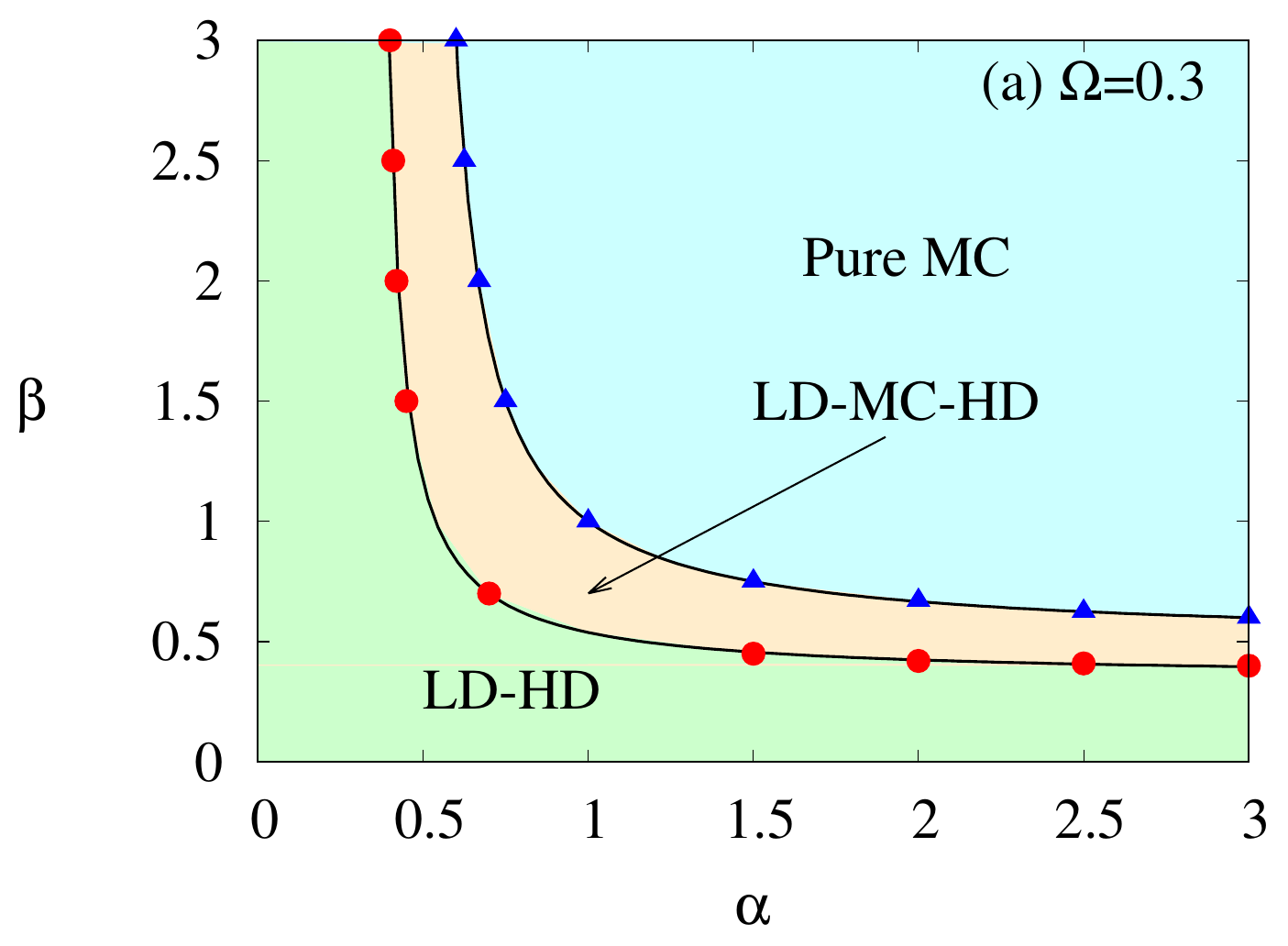}
\label{alpha-vs-beta-Omega-0.3}
\end{minipage}%
\hfill
\begin{minipage}{0.32\linewidth}
\includegraphics[width=\textwidth]{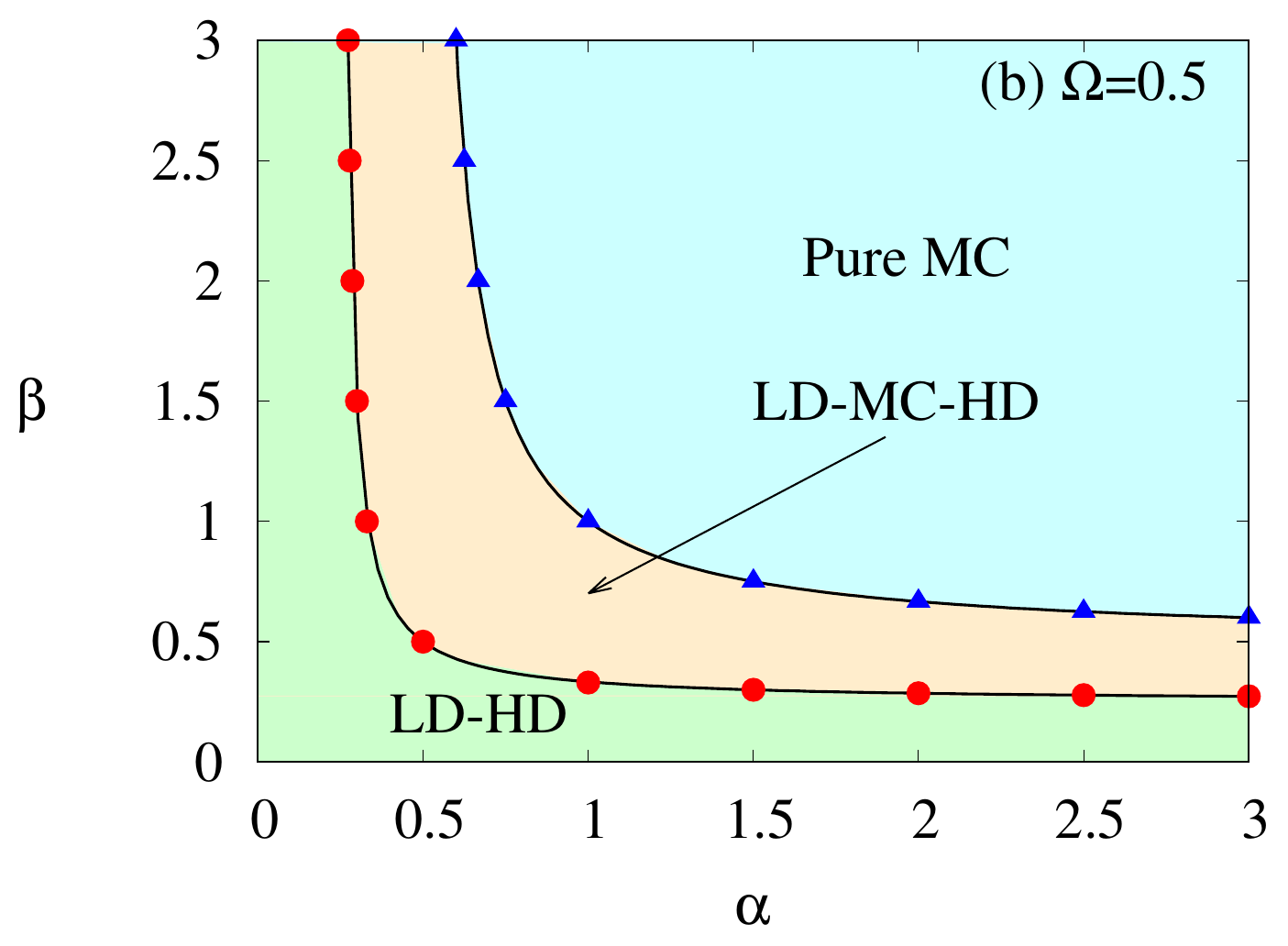}
\label{alpha-vs-beta-Omega-0.5}
\end{minipage}%
\hfill
\begin{minipage}{0.32\linewidth}
\includegraphics[width=\textwidth]{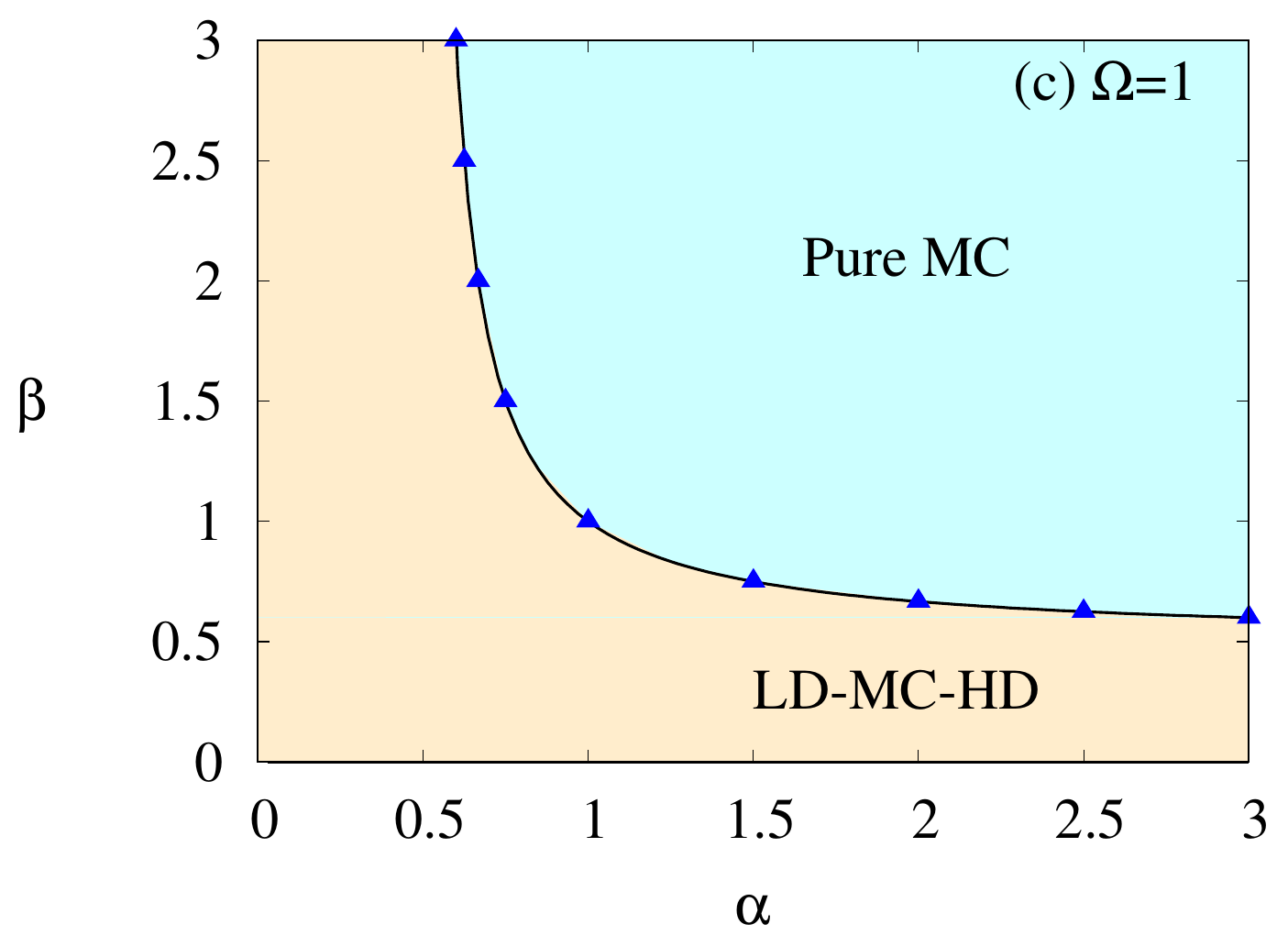}
\label{alpha-vs-beta-Omega-1}
\end{minipage}
\begin{minipage}{0.32\linewidth}
\includegraphics[width=\textwidth]{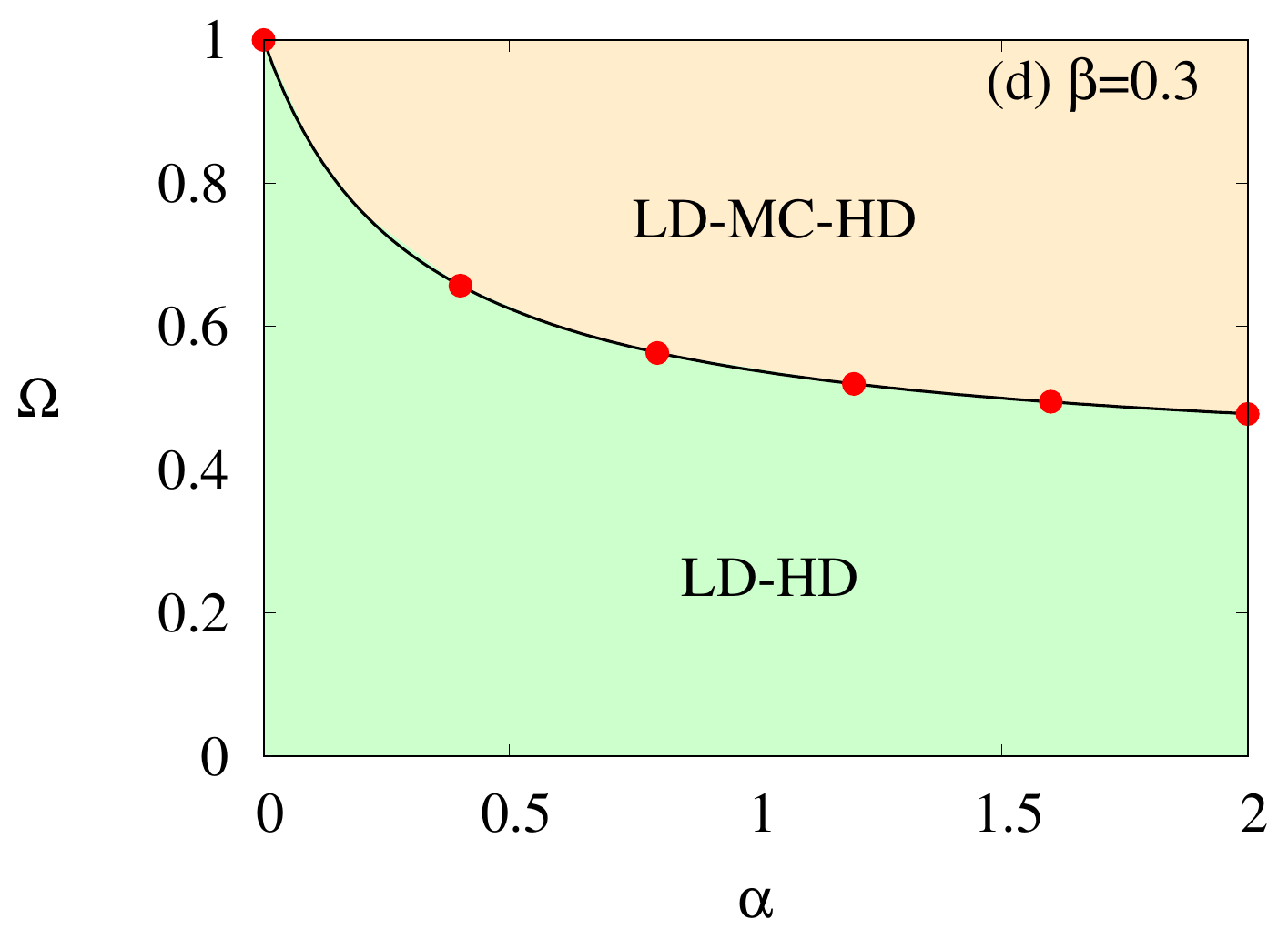}
\label{alpha-vs-Omega-beta-0.3}
\end{minipage}%
\hfill
\begin{minipage}{0.32\linewidth}
\includegraphics[width=\textwidth]{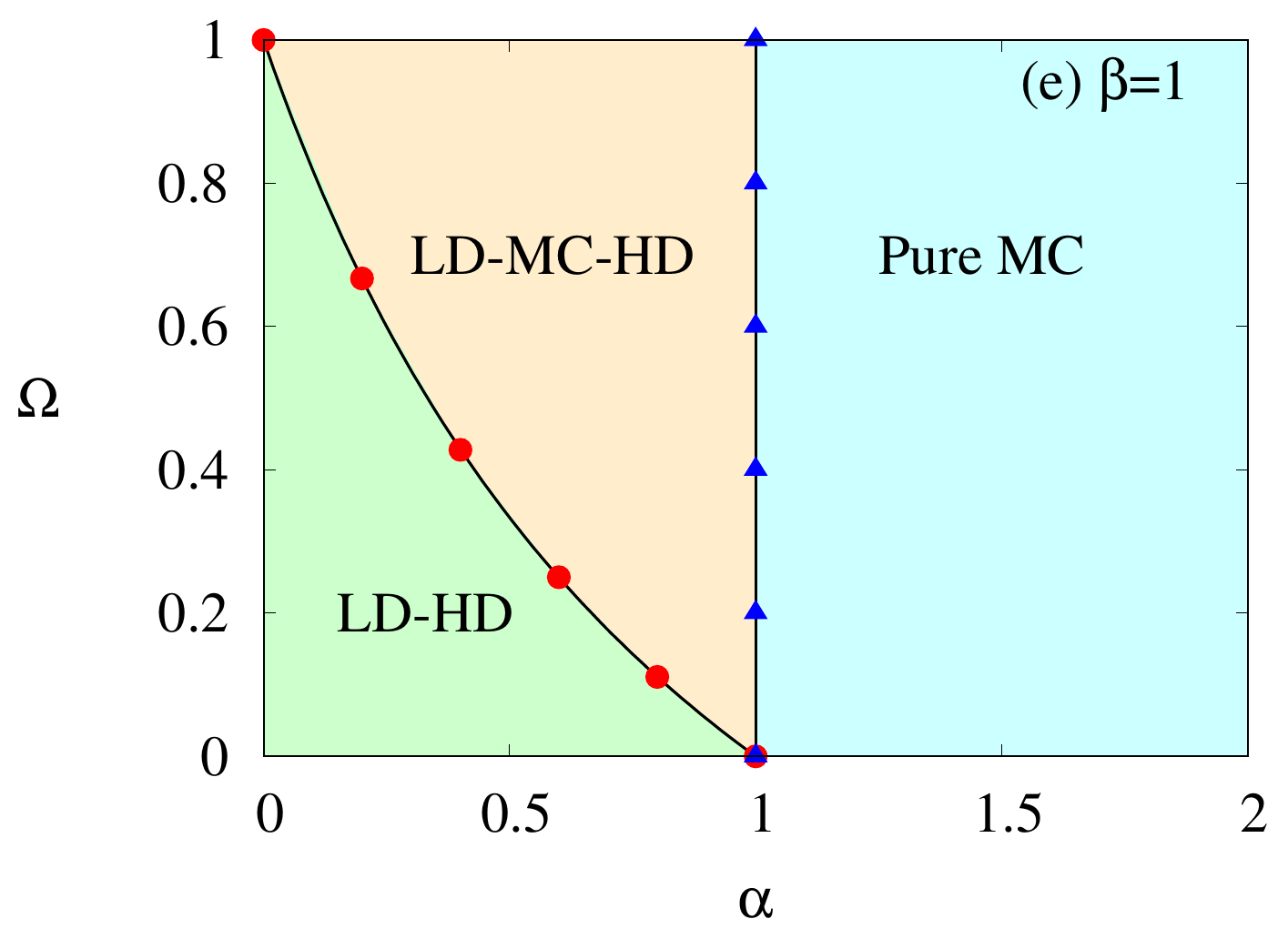}
\label{alpha-vs-Omega-beta-1}
\end{minipage}%
\hfill
\begin{minipage}{0.32\linewidth}
\includegraphics[width=\textwidth]{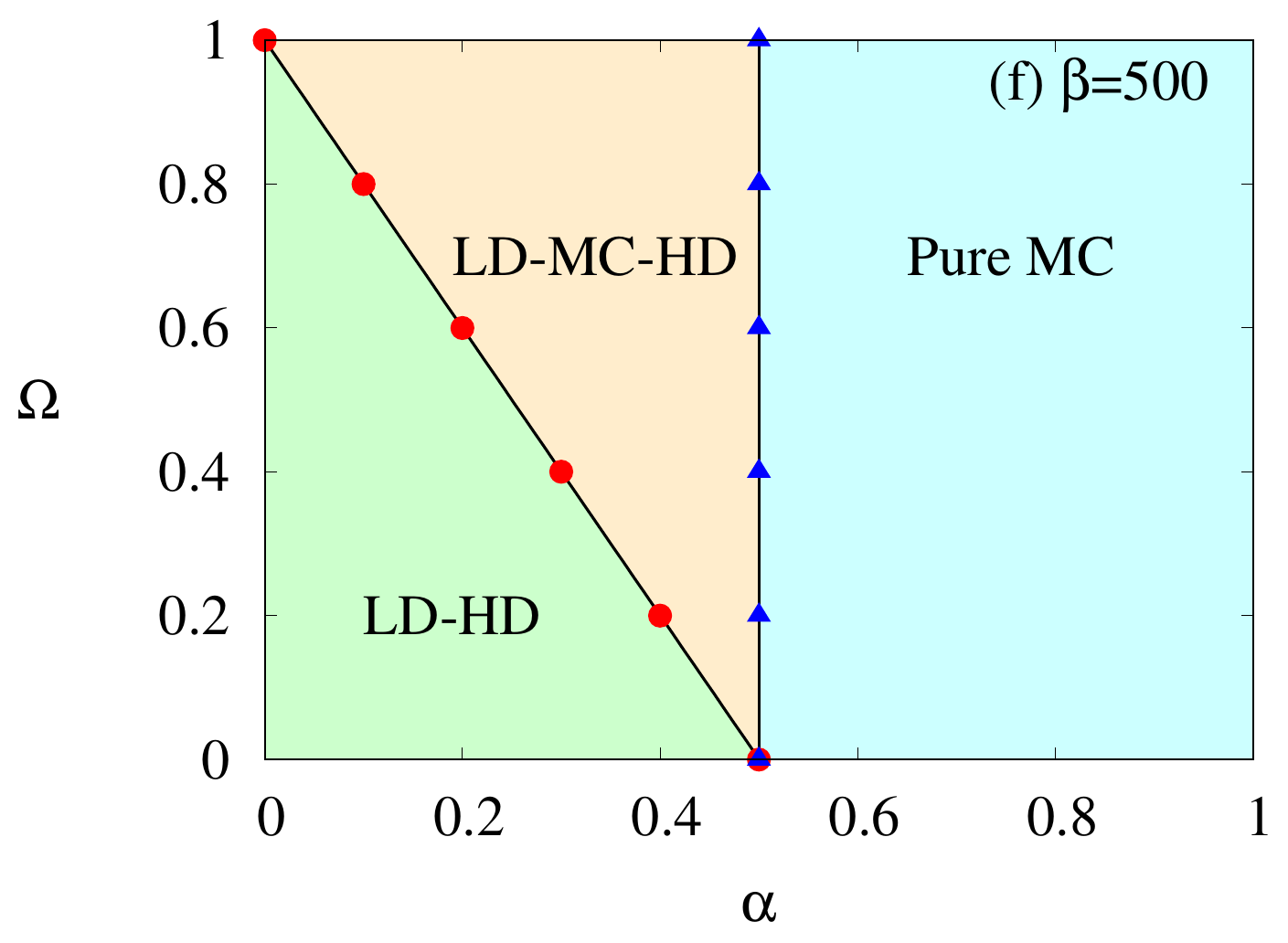}
\label{alpha-vs-Omega-beta-500}
\end{minipage}
\caption{2D phase diagrams of our model are presented. The top panel illustrates the phase diagrams in the $(\alpha, \beta)$-plane for three different values of $\Omega$: (a) $\Omega = 0.3$, (b) $\Omega = 0.5$, and (c) $\Omega = 1$. The bottom panel shows the phase diagrams in the $(\alpha, \Omega)$-plane for three distinct values of $\beta$: (d) $\beta = 0.3$, (e) $\beta = 1$, and (f) $\beta = 500$. The black solid lines represent the phase boundaries obtained from the mean-field analysis, while the colored points correspond to the phase boundaries determined through Monte Carlo simulations. Both results show excellent agreement with each other.
}
\label{2d-pd}
\end{figure}

\end{widetext}

\begin{figure}[!h]
 \centering
\includegraphics[width=\linewidth]{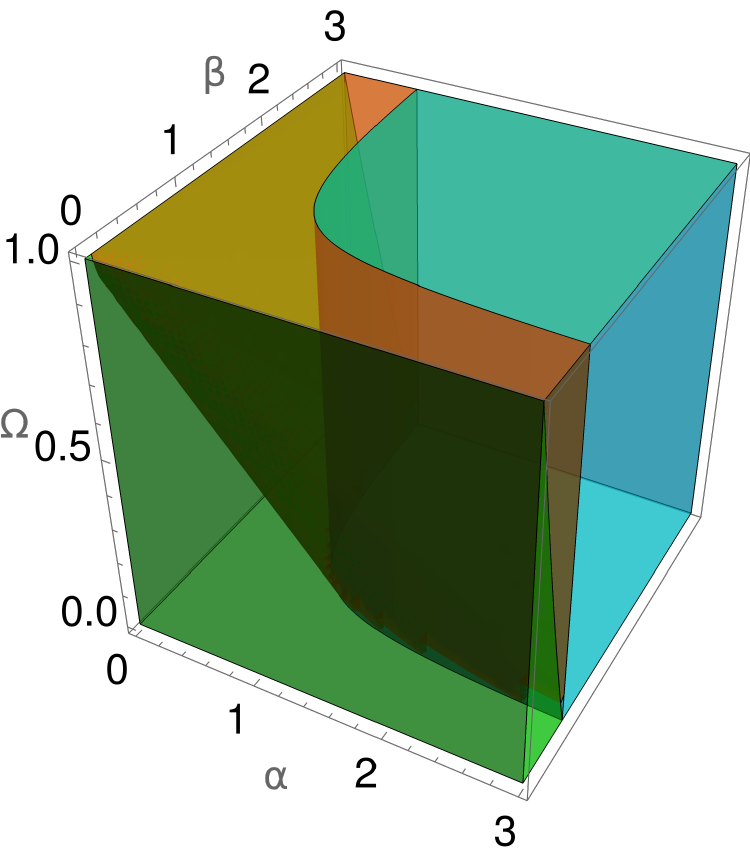}
\caption{Phase diagram in the $\alpha$-$\beta$-$\Omega$-plane exhibits three distinct phases: LD-HD (green), LD-MC-HD (orange), and MC (cyan).}
\label{3d-pd-abomega}
\end{figure}

In this section, we explore the phase diagrams of our model in the space of the control parameters
$\alpha$, $\beta$, and $\Omega$. 

Before turning to our model, it is instructive to briefly revisit the results of
Ref.~\cite{ef-lktasep-pre}, which studied the steady-state properties of an open-boundary TASEP
supplemented with Langmuir kinetics (Lk) for (scaled) equal attachment and detachment rates $\Omega$.
In that model, depending on the entry rate $\alpha$, the exit rate $\beta$,
and the parameter $\Omega$, the stationary density profiles exhibit low-density (LD), high-density (HD),
and maximal-current (MC) phases, as well as coexistence phases involving two or three phases,
such as LD-HD, LD-MC, MC-HD, and LD-MC-HD. By tuning these control parameters, the system undergoes continuous transitions between
different steady states.
In the limit $\Omega \to 0$, the phase diagram of the standard open-boundary TASEP without Langmuir kinetics is recovered.
In contrast, in the limit $\Omega \to \infty$, the Langmuir isotherm
$\rho_{l}=1/2$ dominates the bulk of the system, confining the LD and HD phases
to narrow regions near the boundaries.

The phase diagrams obtained in our model differ markedly from those reported in
Ref.~\cite{ef-lktasep-pre}. Fig.~\ref{2d-pd} shows representative two-dimensional phase diagrams in the $\alpha$-$\beta$ plane for fixed values of $\Omega$, while Fig.~\ref{3d-pd-abomega} presents the corresponding three-dimensional phase diagram in the $\alpha$-$\beta$-$\Omega$ space. 
Although our model exhibits both two-phase and three-phase coexistence regions—namely, the LD–HD and LD–MC–HD phases—as well as a maximal-current (MC) phase, it does not support the LD, HD, LD–MC, and MC–HD phases that are present in Ref.~\cite{ef-lktasep-pre}. In the particle-conserving limit $\Omega=0$, the $\alpha$-$\beta$ plane is occupied only by the LD-HD and MC phases. These are separated by the hyperbola $\alpha\beta/(\alpha+\beta)=1/2$, with the LD-HD phase occurring for $\alpha\beta/(\alpha+\beta)<1/2$ and the MC phase for $\alpha\beta/(\alpha+\beta)>1/2$. The LD-MC-HD phase, defined by the region $(1-\Omega)/2<\alpha\beta/(\alpha+\beta)<1/2$, is therefore absent at $\Omega=0$. As $\Omega$ is increased slightly above zero, the LD-MC-HD phase begins to emerge. With increasing $\Omega$, this three-phase coexistence region expands progressively, accompanied by a corresponding shrinkage of the LD-HD phase, while the MC phase remains unaffected. This evolution continues up to $\Omega=1$, at which point the region $\alpha\beta/(\alpha+\beta)<1/2$ is entirely occupied by the LD-MC-HD phase. Consequently, at $\Omega=1$ the $\alpha$-$\beta$ plane is divided into only two phases: the LD-MC-HD phase for $\alpha\beta/(\alpha+\beta)<1/2$ and the MC phase for $\alpha\beta/(\alpha+\beta)>1/2$, with the LD-HD phase being completely absent. Beyond $\Omega=1$, the overall topology of the phase diagram does not undergo any qualitative change. In the limit $\Omega\to\infty$, the bulk density is governed by the Langmuir isotherm $\rho_l=1/2$, while the LD and HD phases are confined to narrow boundary layers near the system edges. All phase transitions in the system are continuous and are determined by equating the steady-state currents of the competing phases.

The phase diagrams are obtained analytically using mean-field theory (MFT) and are corroborated by extensive Monte Carlo simulations (MCS). The phase boundaries extracted from MCS show excellent agreement with the corresponding MFT predictions.

 \section{MEAN-FIELD ANALYSIS OF THE STEADY STATES}
 \label{mft}

 In this section, we develop a mean-field theory (MFT)~\cite{blythe} to analyze the steady-state density and current profiles, as well as the resulting phase diagrams. Our approach closely follows the reasoning in Ref.~\cite{ef-lktasep-pre}.

Under the MFT approximation, spatial correlations between sites are neglected by factorizing joint occupation probabilities into products of single-site averages:
\begin{equation}
    \langle n_j n_k \rangle \approx \langle n_j \rangle \langle n_k \rangle,
\end{equation}
where $n_j \in \{0,1\}$ denotes the occupation number of site $j$. In the steady state, the particle density at site $j$ is defined as the long-time average of the corresponding occupation variable, $\rho_j = \langle n_j \rangle$. Within the mean-field approximation, densities of the bulk sites $1<j<L$ obey the following equations of motion:
\begin{equation}
\frac{d\rho_j}{dt} \;=\;
\rho_{j-1}(1-\rho_j) - \rho_j(1-\rho_{j+1})
+ \omega(1-\rho_j) - \omega\rho_j,
\label{eom-bulk-sites}
\end{equation}
whereas, the boundary sites $j=1$ and $j=L$ follows
\begin{align}
\frac{d\rho_1}{dt} &= \alpha_{\mathrm{eff}}(1-\rho_1) - \rho_1(1-\rho_2),
\label{eom-site1} \\[6pt]
\frac{d\rho_L}{dt} &= \rho_{L-1}(1-\rho_L) - \beta_{\mathrm{eff}}\rho_L.
\label{eom-siteL}
\end{align}

The equations of motion~\eqref{eom-bulk-sites}--\eqref{eom-siteL} respect particle--hole symmetry under the following transformations:
\begin{eqnarray}
\rho_{j} &\leftrightarrow& 1-\rho_{L-j+1}, \label{tr-den} \\[1mm]
\alpha_{\mathrm{eff}} &\leftrightarrow& \beta_{\mathrm{eff}}, \label{tr-rates}
\end{eqnarray}

Identifying the local hopping current from site $j$ to $j+1$ as $J_j = \rho_j(1-\rho_{j+1})$, the first two terms in Eq.~\eqref{eom-bulk-sites} may be written in divergence form as
\begin{equation}
\rho_{j-1}(1-\rho_j) - \rho_j(1-\rho_{j+1}) = J_{j-1} - J_j, \nonumber
\end{equation}
corresponding to the conservative part of the dynamics, where particles are only transported between neighbouring sites. In contrast, the last two terms in Eq.~\eqref{eom-bulk-sites}, namely $\omega(1-\rho_j)$ and $\omega\rho_j$, arise from Langmuir kinetics and act as source and sink contributions, respectively, constitutes the nonconservating part in the bulk. In the steady state,
\[
\frac{d\rho_j}{dt} = 0 \qquad \text{for } 1 \le j \le L.
\]

To solve the $L$ coupled nonlinear equations~\eqref{eom-bulk-sites}--\eqref{eom-siteL}, we take the continuum limit by coarse-graining the lattice. Introducing the lattice spacing $\epsilon = 1/L$, where $L$ is the system size, we define the rescaled spatial coordinate
\[
x = j\,\epsilon \in [0,1].
\]
In the limit $L \gg 1$, the variable $x$ becomes quasi-continuous, and the discrete density profile $\rho_j$ can be approximated by a smooth function $\rho(x)$.

In the continuum limit, and after rescaling the attachment--detachment rates as in Eq.~\eqref{attachment-detachment-rates}, the steady-state discrete bulk equation~\eqref{eom-bulk-sites} reduces to
\begin{equation}
\label{eom-bulk-cont}
(1 - 2\rho)\,\bigl(\partial_x \rho - \Omega \bigr) = 0,
\end{equation}
where the result holds up to terms of order $\mathcal{O}(\epsilon)$. Note that Eq.~\eqref{eom-bulk-cont} is symmetric under particle-hole exchange, i.e., it is invariant under the transformation~$\rho(x) \leftrightarrow 1-\rho(1-x)$. Unlike the constant boundary conditions in the standard open TASEP with Langmuir kinetics~\cite{ef-lktasep-pre}, the finite reservoir--TASEP coupling in our model leads to dynamically adjusted boundary conditions,
\begin{eqnarray}
\rho(0) &=& \alpha_{\mathrm{eff}}, \hspace{12mm} \text{(left boundary)}, \label{left-bc}\\[4pt]
\rho(1) &=& 1 - \beta_{\mathrm{eff}}, \hspace{7mm} \text{(right boundary)}, \label{right-bc}
\end{eqnarray}
where the effective entry and exit rates ($\alpha_{\mathrm{eff}}$ and $\beta_{\mathrm{eff}}$) introduced in Eq.~\eqref{effective-entry-and-exit-rates} depend on the instantaneous reservoir occupancy through the reservoir feedback function $f$ defined in Eq.~\eqref{f}. Equation~\eqref{eom-bulk-cont} admits three possible steady-state density profiles:
\begin{equation}
\label{den-sol}
\rho(x)=
\begin{cases}
    \rho_{l} = \tfrac{1}{2},\\
    \rho_{\alpha}(x) = \Omega x + \alpha_{\mathrm{eff}},\\
    \rho_{\beta}(x) = \Omega(x-1) + 1 - \beta_{\mathrm{eff}},
\end{cases}
\end{equation}
where $\rho_{l}$ corresponds to the Langmuir isotherm $\rho_{l}=1/2$ with equal attachment and detachment rates, and also represents the maximal-current (MC) density. The density profiles $\rho_{\alpha}(x)$ and $\rho_{\beta}(x)$ are obtained by imposing the left and right boundary conditions, given in Eqs.~\eqref{left-bc} and~\eqref{right-bc}, respectively, on the linear bulk solution $\rho(x) = \Omega x + \text{const.}$ of Eq.~\eqref{eom-bulk-cont}.

Within mean-field theory, in the thermodynamic limit ($L \to \infty$ or equivalently $\epsilon \to 0$), the stationary particle current corresponding to a given density profile is, to leading order,
\begin{equation}
\label{mf-current}
J(x) = \rho(x)\,\bigl(1-\rho(x)\bigr).
\end{equation}
Unlike systems with purely TASEP dynamics (i.e., without Langmuir kinetics), the current profiles here vary smoothly with the spatial coordinate $x$.

The time evolution of the reservoir population $N_R$ is governed by
\begin{equation}
\label{NR-time-evolution}
\frac{dN_R}{dt} = -\rho(0)\bigl(1-\rho(0)\bigr) + \rho(1)\bigl(1-\rho(1)\bigr).
\end{equation}
In the steady state, $dN_R/dt = 0$. Using Eqs.~\eqref{effective-entry-and-exit-rates}, \eqref{left-bc}, and \eqref{right-bc} in Eq.~\eqref{NR-time-evolution}, one obtains a quadratic equation in $f(N_{R})$ for $\alpha \neq \beta$, which has following solutions:
\begin{equation}
\label{f-plus-minus}
f_{\pm}(\alpha,\beta) = \frac{\alpha+\beta-2\beta^2}{2(\alpha^2-\beta^2)} \pm
\sqrt{\left(\frac{\alpha+\beta-2\beta^2}{2(\alpha^2-\beta^2)}\right)^2 - \frac{\beta(1-\beta)}{\alpha^2-\beta^2}}.
\end{equation}
Since $0 \le f(N_R) \le 1$, only $f_-(N_R)$ is the physically acceptable solution. After simplification, it reduces to
\begin{equation}
\label{f-minus}
f_-(\alpha,\beta) = \frac{\beta}{\alpha+\beta}.
\end{equation}
Substituting Eq.~\eqref{f-minus} into Eq.~\eqref{effective-entry-and-exit-rates} gives the effective entry and exit rates:
\begin{equation}
\label{alpha-beta-eff-same}
\alpha_{\mathrm{eff}} = \beta_{\mathrm{eff}} = \frac{\alpha \beta}{\alpha+\beta}.
\end{equation}
Substituting Eq.~\eqref{alpha-beta-eff-same} in Eq.~\eqref{den-sol}, one gets
\begin{equation}
\label{den-sol-cont-par}
\rho(x)=
\begin{cases}
    \rho_{l} = \tfrac{1}{2},\\
    \rho_{\alpha}(x) = \Omega x + \frac{\alpha \beta}{\alpha+\beta},\\
    \rho_{\beta}(x) = \Omega(x-1) + 1 - \frac{\alpha \beta}{\alpha+\beta},
\end{cases}
\end{equation}

Having obtained the density solutions in Eq.~(\ref{den-sol-cont-par}) in terms of the control parameters $\alpha$, $\beta$, and $\Omega$, we now construct the full density profiles. The full density profiles emerge from different scenarios depending on how the three solutions, $\rho_l$, $\rho_\alpha(x)$, and $\rho_\beta(x)$, are matched.

In the following, we consider two cases that define two distinct parameter regimes based on the effective boundary densities defined in Eq.~\eqref{alpha-beta-eff-same}:
\begin{itemize}
    \item \textit{Case~I:} $\alpha_{\mathrm{eff}}=\beta_{\mathrm{eff}}=\dfrac{\alpha\beta}{\alpha+\beta}<\dfrac{1}{2}$.
    \item \textit{Case~II:} $\alpha_{\mathrm{eff}}=\beta_{\mathrm{eff}}=\dfrac{\alpha\beta}{\alpha+\beta}>\dfrac{1}{2}$.
\end{itemize}

 Let us first consider Case~I:
\begin{equation}
\label{mc-cond}
\alpha_{\mathrm{eff}}=\beta_{\mathrm{eff}}=\frac{\alpha \beta}{\alpha+\beta}<\frac{1}{2}.
\end{equation}
Under this condition, the boundary densities~\eqref{left-bc} and~\eqref{right-bc} turn out to be $\rho(0)<1/2$ and $\rho(1)>1/2$. Consequently, the left solution $\rho_{\alpha}(x)$, which starts from $\rho_{\alpha}(0)=\alpha_{\mathrm{eff}}<1/2$, increases linearly with slope $\Omega$ and intersects the constant solution $\rho_l=1/2$ at some position $x_\alpha$. Using Eq.~\eqref{den-sol-cont-par}, this intersection point is determined from
\begin{eqnarray}
\rho_{\alpha}(x_{\alpha}) = \Omega x_{\alpha} + \frac{\alpha \beta}{\alpha + \beta} = \frac{1}{2}, \nonumber \\
\implies x_{\alpha} = \frac{1 - \frac{2\alpha \beta}{\alpha + \beta}}{2\Omega}.
\label{xa}
\end{eqnarray}
Similarly, the right solution $\rho_{\beta}(x)$ originates from $\rho_{\beta}(1)=1-\beta_{\mathrm{eff}}>1/2$, and decreases linearly with slope $\Omega$ as $x$ moves leftwards. This branch intersects the constant solution $\rho_{l}=1/2$ at a position $x_{\beta}$, which follows from Eq.~\eqref{den-sol-cont-par} as
\begin{eqnarray}
\rho_{\beta}(x_{\beta})=\Omega (x_{\beta}-1)+1-\frac{\alpha \beta}{\alpha+\beta}=\frac{1}{2}, \nonumber \\
\implies x_{\beta}=\frac{2\Omega+\frac{2\alpha\beta}{\alpha+\beta}-1}{2\Omega}.
\label{xb}
\end{eqnarray}
It is evident from Eqs.~(\ref{xa}) and (\ref{xb}) that
$x_{\alpha} \rightarrow 0^{+}$ as
$\alpha_{\text{eff}} = \alpha\beta/(\alpha+\beta) \rightarrow 1/2^{-}$
and
$x_{\beta} \rightarrow 1^{-}$ as
$\beta_{\text{eff}} = \alpha\beta/(\alpha+\beta) \rightarrow 1/2^{-}$.
Three scenarios may emerge depending on the relative values of
$x_{\alpha}$ and $x_{\beta}$:
 \\

  \vspace{1mm}

  (i) $x_{\alpha}>x_{\beta}$:
Using Eqs.~(\ref{xa}) and (\ref{xb}), this condition translates to
\begin{equation}
 \frac{\alpha\beta}{\alpha+\beta} < \frac{1-\Omega}{2}.
 \label{dw-cond}
\end{equation}

In this regime, the density profile contains a domain wall (DW) within the system boundaries. The stationary density profile is then composed of two linear branches: the low-density (LD) solution $\rho_{\alpha}(x)$ originating from the left boundary, and the high-density (HD) solution $\rho_{\beta}(x)$ originating from the right boundary. These meet at a domain wall located at position $x_{w}$, giving a piecewise form
\begin{equation}
\label{xa-greater-xb}
\rho(x)=
\begin{cases}
\rho_{\alpha}(x) = \Omega x + \frac{\alpha\beta}{\alpha+\beta},  & 0 \le x \le x_{w}, \\[2mm]
\rho_{\beta}(x) = \Omega(x-1) + 1 - \frac{\alpha\beta}{\alpha+\beta}, & x_{w} \le x \le 1.
\end{cases}
\end{equation}

This corresponds to a two–phase coexistence region:
the segment $0 \le x \le x_{w}$ remains in a LD phase with
$\rho_\alpha(x) < 1/2$ and $J_\alpha(x) = \rho_\alpha(x)(1-\rho_\alpha(x)) < 1/4$, while the segment $x_{w} \le x \le 1$ remains in a HD phase with $\rho_\beta(x) > 1/2$ and $J_\beta(x) = \rho_\beta(x)(1-\rho_\beta(x)) < 1/4$.

The position of the domain wall is determined by enforcing continuity of the stationary current between the LD and HD branches, i.e.,
\begin{equation}
\label{equal-current}
J_{\alpha}(x_w)=J_{\beta}(x_w),
\end{equation}
where the current $J(x)$ is given by Eq.~\eqref{mf-current}. Substituting the LD and HD profiles from Eq.~\eqref{xa-greater-xb} into Eq.~\eqref{equal-current} yields two possible solutions for the matching point $x_w$:
\begin{equation}
\label{dw-unphysical}
\rho_{\alpha}(x_w)=\rho_{\beta}(x_w),
\end{equation}
or
\begin{equation}
\label{dw-den-cond}
\rho_{\alpha}(x_w)+\rho_{\beta}(x_w)=1.
\end{equation}
Since a domain wall corresponds to a finite discontinuity in the density, Eq.~\eqref{dw-unphysical} must be discarded. Enforcing the physically relevant solution~\eqref{dw-den-cond}, one finds
\begin{equation}
\label{dw-pos}
x_w=\frac{1}{2}.
\end{equation}
Thus, within this LD-HD coexistence regime, the domain wall remains \textit{strictly localized} at the center of the system, independent of the control parameters $\alpha$, $\beta$, and $\Omega$. This is in stark contrast to the result obtained in Ref.~\cite{ef-lktasep-pre} where although the domain wall is localized, its position depends on the control parameters of the model.

Finally, the height $\mathcal{H}$ of the domain wall is
\begin{align}
\label{dw-height}
\begin{split}
\mathcal{H} &= \rho_{\beta}(x_{w}) - \rho_{\alpha}(x_{w}) \\[1mm]
  &= 1 - \Omega - \frac{2\alpha\beta}{\alpha+\beta}.
\end{split}
\end{align}

\begin{figure*}[ht]
\includegraphics[width=\textwidth]{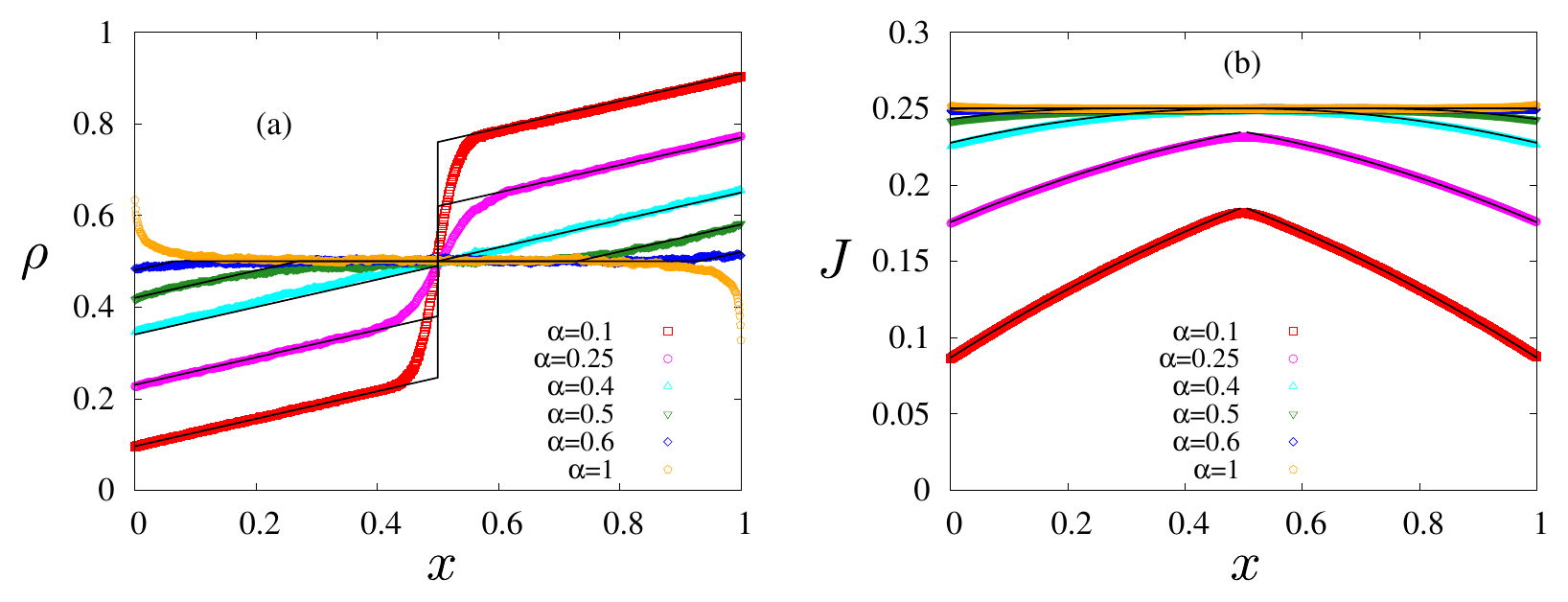}
\caption{(a) Density and (b) current profiles for different values of $\alpha$ with fixed $\beta = 2.5$ and $\Omega = 0.3$. 
For $\alpha = 0.1,\,0.25,\,0.4$, the system is in the LD-HD phase with density given by Eq.~\eqref{xa-greater-xb} and a domain wall located at $x_{w} = 1/2$ (Eq.~\eqref{dw-pos}). As $\alpha$ increases, the domain-wall height (Eq.~\eqref{dw-height}) decreases continuously and vanishes at $\alpha = 0.4$, producing a fully linear profile. 
For $\alpha = 0.5,\,0.6$, the system enters the LD-MC-HD phase described by Eq.~\eqref{xa-less-xb}, where the width of the maximal-current region (Eq.~\eqref{mc-phase-width}) grows with increasing $\alpha$. At $\alpha = 1$, the system reaches the MC phase with bulk density $1/2$ and boundary layers of vanishing width in the thermodynamic limit. 
The density changes smoothly across LD-HD/LD-MC-HD and LD-MC-HD/MC boundaries, indicating second-order transitions. The corresponding current profiles in Fig.~\ref{den-curr-profile1}(b) also vary smoothly with $x$. 
Simulations are performed for system size $L = 1000$, with averages taken over $10^{6}$ Monte Carlo steps. The Monte Carlo results (colored points) show excellent agreement with the mean-field predictions (black solid lines).
}
\label{den-curr-profile1} 
\end{figure*}

\begin{figure*}[ht]
\includegraphics[width=\textwidth]{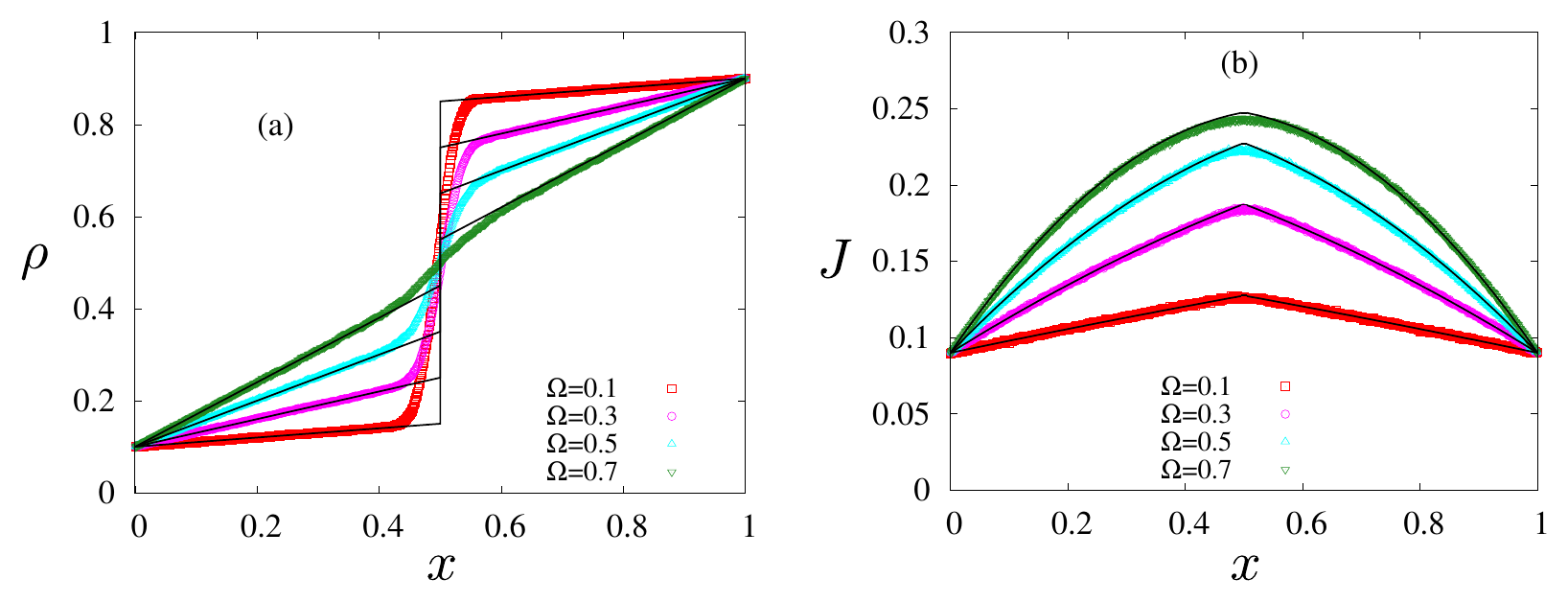}
\caption{(a) Density and (b) current profiles in the LD-HD phase for different values of $\Omega$ with fixed $\alpha = \beta = 0.2$. The density profiles (Eq.~\eqref{xa-greater-xb}) show that the slope increases with increasing $\Omega$, whereas the height of the DW (Eq.~\eqref{dw-height}) decreases with increasing $\Omega$ for a fixed value of $\alpha,\,\beta$. The corresponding current profiles are shown in Fig.~\ref{den-curr-profile2}(b). 
Simulations are performed for a system size of $L = 1000$, with time-averaging over $10^{6}$ Monte Carlo steps. The Monte Carlo data (colored points) show excellent agreement with the mean-field predictions (black solid lines).}
\label{den-curr-profile2} 
\end{figure*}

The corresponding density and current profiles are shown in
Figs.~\ref{den-curr-profile1}(a,b) for $\alpha=0.1,\,0.25$ at fixed
$\beta=2.5$ and $\Omega=0.3$, and in Figs.~\ref{den-curr-profile2}(a,b)
for fixed $\alpha=\beta=0.2$ and varying
$\Omega=0.1,\,0.3,\,0.5,$ and $0.7$.
The density profiles show that the solutions
$\rho_{\alpha}(x)$ and $\rho_{\beta}(x)$ vary linearly with $x$ and
exhibit a discontinuity at the domain-wall position $x_w=1/2$. The density profiles [Eq.~\eqref{xa-greater-xb}] show that, for fixed values of $\alpha$ and $\beta$, the slope increases with increasing $\Omega$, while the height of the domain wall [Eq.~\eqref{dw-height}] decreases as $\Omega$ increases; see Fig.~\ref{den-curr-profile2}(a). The corresponding current profiles demonstrate that
$J_{\alpha}(x)$ increases monotonically with $x$, whereas
$J_{\beta}(x)$ decreases monotonically, with both branches meeting at
$x_w=1/2$ indicating a localized domain wall.

  \vspace{1mm}

  (ii)~$x_{\alpha}=x_{\beta}$: Substituting this condition into Eqs.~\eqref{xa} and \eqref{xb} yields the hyperbola:
\begin{equation}
\label{xa-equals-xb}
\frac{\alpha\beta}{\alpha+\beta} = \frac{1-\Omega}{2}.
\end{equation}
Approaching this curve from the LD-HD coexistence region, where
\[
\frac{\alpha\beta}{\alpha+\beta} < \frac{1-\Omega}{2},
\]
the height of the domain wall (see Eq.~\eqref{dw-height}), located at $x_w=1/2$, decreases smoothly and eventually vanishes at the hyperbola. For parameters on this line, the density profile becomes fully continuous, forming a single linear branch connecting the LD and HD boundary conditions across the midpoint $x=1/2$. The corresponding steady-state density profile is given by 
\begin{equation}
\label{continuous-profile-hyperbola}
\rho(x) = \Omega x + \frac{1-\Omega}{2}, \qquad 0 \le x \le 1,
\end{equation}
which can be obtained from the LD-HD density profile in Eq.~\eqref{xa-greater-xb} by substituting Eq.~\eqref{xa-equals-xb}. The current profile is given by
\begin{equation}
J(x) = \left(\Omega x \;+\; \frac{1-\Omega}{2}\right)
\left( 1 - \Omega x - \frac{1-\Omega}{2} \right),
\hfill 0 \le x \le 1 ,
\label{current-cont-profile}
\end{equation}
Representative density profile is shown in Fig.~\ref{den-curr-profile1}(a) for $\alpha=0.4$, $\beta=2.5$, and $\Omega=0.3$, along with the current profile in Fig.~\ref{den-curr-profile1}(b).

\begin{figure*}[ht]
\includegraphics[width=\textwidth]{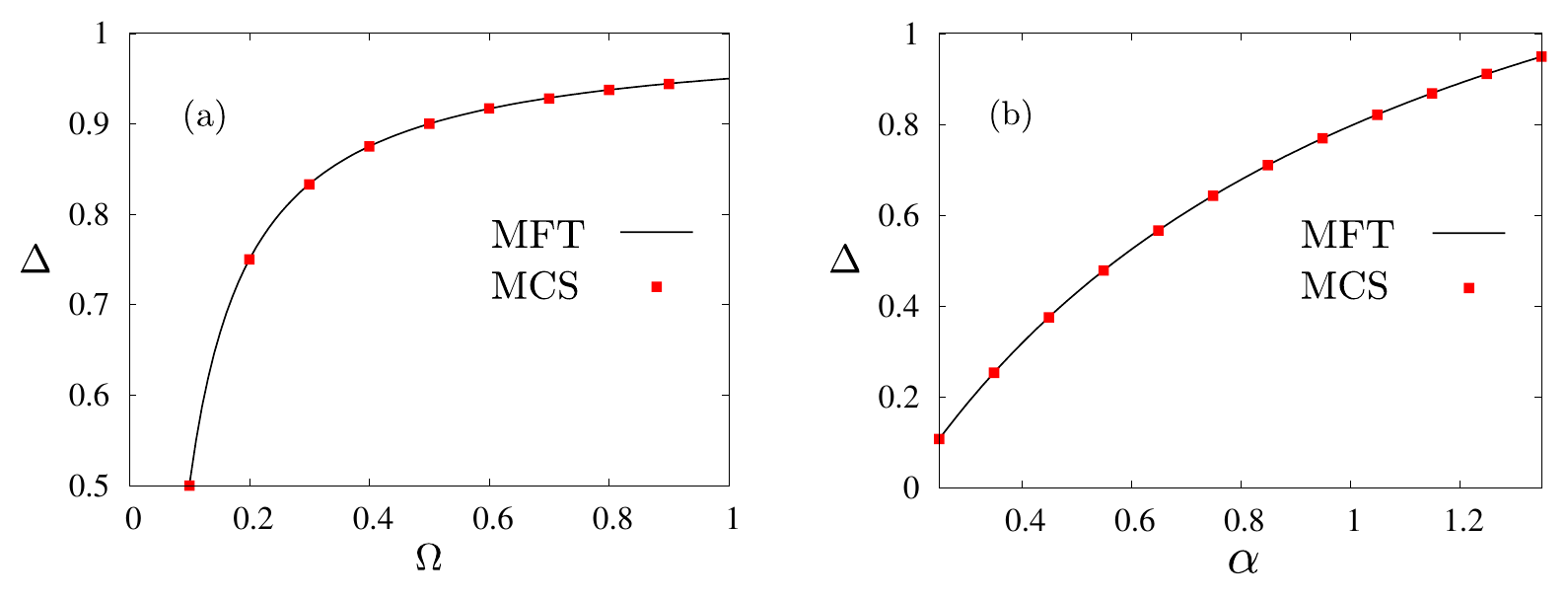}
\caption{Plots of the width $\Delta$ of the MC phase segment within the LD-MC-HD phase are shown as functions of (a) $\Omega$ for fixed $\alpha=\beta=0.95$, and (b) $\alpha$ for fixed $\beta=0.75$ and $\Omega=0.7$. The mean-field theory (MFT) predictions (black solid lines) show excellent agreement with the corresponding Monte Carlo simulation (MCS) results (colored symbols). Evidently, $\Delta$ increases monotonically with $\Omega$ for fixed $\alpha$ and $\beta$, and also increases with $\alpha$ for fixed $\beta$ and $\Omega$, in accordance with Eq.~\eqref{mc-phase-width}.}
\label{del-vs-omega-alpha-plots} 
\end{figure*}

\begin{figure}[!h]
 \centering
 \includegraphics[width=\columnwidth]{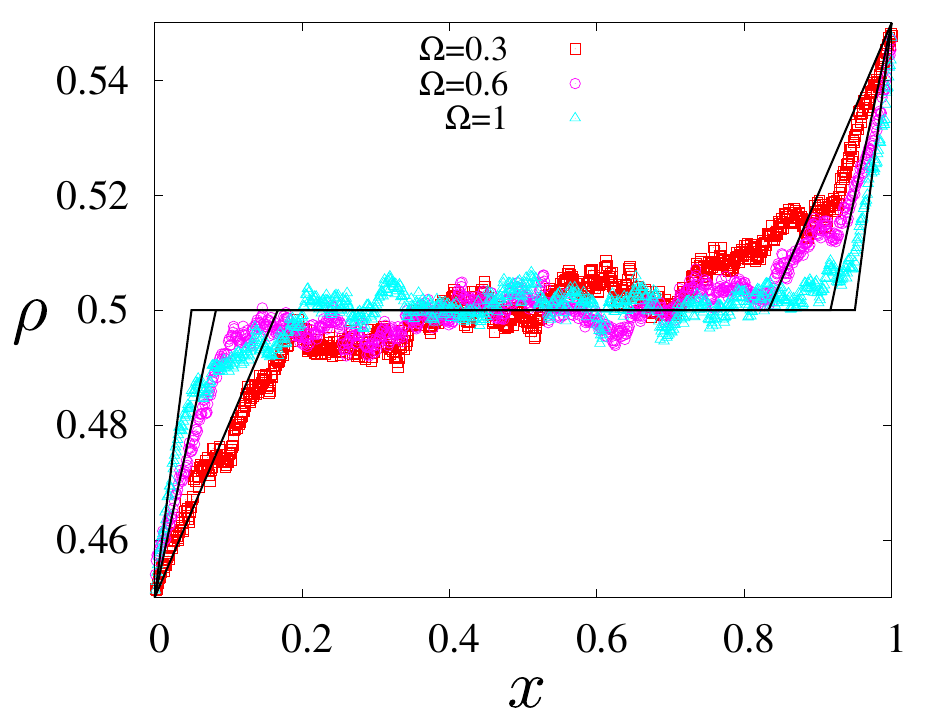}
 \caption{Density profiles in the LD-MC-HD phase for fixed $\alpha=\beta=0.9$ and three different values of $\Omega$. The width of the MC region [see Eq.~\eqref{mc-phase-width}] increases with increasing $\Omega$ at fixed $\alpha$ and $\beta$. Colored symbols represent MCS results, which are in excellent agreement with the solid black curves obtained from MFT.}
 \label{mc-part-width}
 \end{figure}

  (iii)~$x_{\alpha}<x_{\beta}$: Substituting this condition in Eqs.~\eqref{xa} and \eqref{xb} gives:
\begin{equation}
\frac{\alpha\beta}{\alpha+\beta} > \frac{1-\Omega}{2}.
\label{ld-mc-hd-phase}
\end{equation}
In this case, the LD and HD branches are separated by a region where the density saturates to its maximal–current value $\rho_\text{MC} = 1/2$. The stationary density profile is continuous and piecewise linear:
\begin{equation}
\label{xa-less-xb}
\rho(x)=
\begin{cases}
\rho_{\alpha}(x) = \Omega x + \frac{\alpha\beta}{\alpha+\beta},
& 0 \le x \le x_{\alpha}, \\[4pt]
\rho_{l} = 1/2,
& x_{\alpha} \le x \le x_{\beta}, \\[4pt]
\rho_{\beta}(x) = \Omega(x-1)+1-\frac{\alpha\beta}{\alpha+\beta},
& x_{\beta} \le x \le 1~.
\end{cases}
\end{equation}
Thus, a three-phase coexistence region emerges: an LD segment with $\rho_\alpha(x)<1/2$ and $J_\alpha(x)=\rho_\alpha(x)(1-\rho_\alpha(x))<1/4$ for $0 \le x \le x_{\alpha}$, a maximal-current (MC) segment with $\rho(x)=1/2$ and $J(x)=1/4$ for $x_{\alpha} \le x \le x_{\beta}$, and an HD segment with $\rho_\beta(x)>1/2$ and $J_\beta(x)=\rho_\beta(x)(1-\rho_\beta(x))<1/4$ for $x_{\beta} \le x \le 1$. The width $\Delta$ of the MC phase segment is given by
\begin{equation}
\label{mc-phase-width}
\Delta = x_{\beta}-x_{\alpha}
= 1 + \frac{1}{\Omega}\left(\frac{2\alpha\beta}{\alpha+\beta} - 1\right).
\end{equation}
The LD-MC-HD phase exists in the parameter regime
\begin{equation}
 \frac{1-\Omega}{2} < \frac{\alpha\beta}{\alpha+\beta} < \frac{1}{2},
 \label{ldmchd-region}
\end{equation}
which implies that the width $\Delta$ [see Eq.~\eqref{mc-phase-width}] increases monotonically with $\Omega$ for fixed $\alpha$ and $\beta$ [see Figs.~\ref{del-vs-omega-alpha-plots}(a) and \ref{mc-part-width}], and also increases with $\alpha$ for fixed $\beta$ and $\Omega$ [see Fig.~\ref{del-vs-omega-alpha-plots}(b)].

As the boundaries of the LD and HD regions move toward the system edges, i.e., $x_{\alpha} \to 0^{+}$ and $x_{\beta} \to 1^{-}$, corresponding to $\alpha_\text{eff}=\beta_\text{eff}=\alpha\beta/(\alpha+\beta) \to 1/2^{-}$, the MC region expands and ultimately spans the entire bulk.

Representative density profiles are shown in Fig.~\ref{den-curr-profile1}(a) for $\alpha=0.5,\,0.6$ with fixed $\beta=2.5$ and $\Omega=0.3$, and the current profiles in Fig.~\ref{den-curr-profile1}(b).

We now analyze Case~II:
\begin{equation}
\label{case-2}
\alpha_{\mathrm{eff}}=\beta_{\mathrm{eff}}=\frac{\alpha \beta}{\alpha+\beta}>\frac{1}{2}.
\end{equation}
This condition implies the boundary values in Eqs.~\eqref{left-bc} and \eqref{right-bc} satisfy $\rho(0)>1/2$ and $\rho(1)<1/2$. In the regime of $\alpha$-$\beta$ plane defined by Eq.~\eqref{case-2}, the system exhibits an MC phase similar to an open-boundary TASEP without Lk dynamics. The MC phase is associated with boundary layers (BLs) at both entry and exit ends which have vanishing thickness in the thermodynamic limit $L \to \infty$. The bulk density is, however, constant on average:
\begin{equation}
 \rho_\text{MC} = \frac{1}{2}. \label{mc-den}
\end{equation}

Density and current profiles for MC phase are shown in Figs.~\ref{den-curr-profile1}(a) and \ref{den-curr-profile1}(b) for parameters $\alpha=1$, $\beta=2.5$, and $\Omega=0.3$.

Table~\ref{tab1} lists the possible phases in the $\alpha$-$\beta$ plane for a fixed $\Omega$ with $0<\Omega<1$.
\begin{widetext}

\begin{table}[h!]
\centering
\renewcommand{\arraystretch}{1.35} 
\setlength{\tabcolsep}{10pt}      
\caption{Possible steady-state phases and their corresponding regions in the $\alpha$-$\beta$ plane for $0 < \Omega < 1$.}
\begin{tabular}{|c|c|}
  \hline
  \textbf{Phase} & \textbf{Region in the $\alpha$-$\beta$ plane} \\
  \hline
  LD-HD (two-phase coexistence) & $\dfrac{\alpha\beta}{\alpha+\beta} < \dfrac{1-\Omega}{2}$ \\
  \hline
  Linear density profile & $\dfrac{\alpha\beta}{\alpha+\beta} = \dfrac{1-\Omega}{2}$ \\
  \hline
  LD-MC-HD (three-phase coexistence) & $\dfrac{1-\Omega}{2} < \dfrac{\alpha\beta}{\alpha+\beta} < \dfrac{1}{2}$ \\
  \hline
  Pure MC & $\dfrac{\alpha\beta}{\alpha+\beta} > \dfrac{1}{2}$ \\
  \hline
\end{tabular}
\label{tab1}
\end{table}

\end{widetext}

\subsection{Absence of LD, HD, LD-MC, and MC-HD phases}
\label{absence}

The phase diagram of our model, shown in Fig.~\ref{2d-pd}, exhibits three distinct steady-state phases: the two-phase coexistence LD--HD, the three-phase coexistence LD--MC--HD, and the homogeneous MC phase. Notably, the phases LD, HD, LD-MC, and MC-HD, which appear in the phase diagrams reported in Ref.~\cite{ef-lktasep-pre}, are absent in our model. In the following, we explain the reason behind the non-feasibility of these phases in the present setting.

\textit{Absence of LD and HD phases:}
In our model, the specific choice of the effective entry and exit rates in Eq.~\eqref{effective-entry-and-exit-rates} and the reservoir feedback function $f(N_{R})$ in Eq.~\eqref{f} results in identical effective entry and exit rates, as given in Eq.~\eqref{alpha-beta-eff-same}. This immediately precludes the existence of a homogeneous LD (or HD) phase, which requires $\alpha_{\text{eff}} < \beta_{\text{eff}}$ (or $\alpha_{\text{eff}} > \beta_{\text{eff}}$).

The realization of an LD phase requires the conditions $x_{\alpha} > 1$ and $x_{\beta} > 1$, where $x_{\alpha}$ and $x_{\beta}$ denote the positions at which the linear density solutions $\rho_{\alpha}(x)$ and $\rho_{\beta}(x)$, originating from the left and right boundaries, respectively, intersect the Langmuir isotherm $\rho_{l} = 1/2$ (see Eq.~\eqref{den-sol-cont-par}). These positions are defined explicitly in Eqs.~\eqref{xa} and~\eqref{xb}. However, the simultaneous conditions $x_{\alpha} > 1$ and $x_{\beta} > 1$ lead to the mutually contradictory inequalities
\begin{eqnarray}
\frac{\alpha \beta}{\alpha + \beta} &<& \frac{1}{2} - \Omega, \quad (\Omega > 0), \label{no-ld-hd-1} \\
\frac{\alpha \beta}{\alpha + \beta} &>& \frac{1}{2}, \label{no-ld-hd-2}
\end{eqnarray}
which cannot be satisfied simultaneously.

A similar argument applies to the HD phase. Its existence requires $x_{\alpha} < 0$ and $x_{\beta} < 0$, which again reduce to the same pair of incompatible conditions, Eqs.~\eqref{no-ld-hd-1} and~\eqref{no-ld-hd-2}. Consequently, both the LD and HD phases are ruled out in the present model.

\textit{Absence of LD-MC and MC-HD phases:}
The LD-MC phase can be obtained as a limit of the LD-MC-HD phase under the conditions
$0 < x_{\alpha} < 1$, $x_{\beta} > 1$, and $x_{\alpha} < x_{\beta}$.
These conditions lead to the following set of inequalities:
\begin{eqnarray}
\frac{\alpha \beta}{\alpha + \beta} &<& \frac{1}{2}, \label{no-ldmc-1} \\
\frac{\alpha \beta}{\alpha + \beta} &>& \frac{1}{2} - \Omega, \label{no-ldmc-2} \\
\frac{\alpha \beta}{\alpha + \beta} &>& \frac{1}{2}, \label{no-ldmc-3} \\
\frac{\alpha \beta}{\alpha + \beta} &>& \frac{1 - \Omega}{2}. \label{no-ldmc-4}
\end{eqnarray}
However, the inequalities~\eqref{no-ldmc-1} and~\eqref{no-ldmc-3} are mutually contradictory and therefore
rule out the existence of the LD-MC phase in our model.

Similarly, the MC-HD phase could arise from the LD-MC-HD phase under the conditions
$x_{\alpha} < 0$, $0 < x_{\beta} < 1$, and $x_{\alpha} < x_{\beta}$.
These conditions lead to the same set of inequalities~\eqref{no-ldmc-1}--\eqref{no-ldmc-4},
which are again inconsistent.
Consequently, the MC-HD phase is also forbidden in our model.

\subsection{Phase boundaries and phase transitions}
\label{pb-pt}

In this section, we determine the boundaries demarcating the phases in our model and characterize the associated transitions.

For reference, see Fig.~\ref{den-curr-profile1}(a), which illustrates the density profiles in all phases for a set of representative values of $\alpha$, with $\beta=2.5$ and $\Omega=0.3$ held fixed. The LD-HD phase, characterized by a density profile in Eq.~\eqref{xa-greater-xb} and a nonzero DW height in Eq.~\eqref{dw-height}, is realized when $x_{\alpha} > x_{\beta}$, where $x_{\alpha}$ and $x_{\beta}$ are the positions where the two linear solutions emerging from the left and right boundaries meet the line $\rho_{l}=1/2$ and are  defined in Eqs.~\eqref{xa} and \eqref{xb}, respectively. When $x_{\alpha} = x_{\beta}$, the DW height vanishes and the density profile becomes a continuous linear function of $x$ given by Eq.~\eqref{continuous-profile-hyperbola}. When $x_{\alpha} < x_{\beta}$, the system transitions into the LD-MC-HD phase. Thus, the boundary between LD-HD and LD-MC-HD phases is determined by the condition $x_{\alpha} = x_{\beta}$, which yields
\begin{equation}
\frac{\alpha \beta}{\alpha + \beta} = \frac{1-\Omega}{2}.
\label{ldhd-and-ldmchd-boundary}
\end{equation}
See the 2D phase diagrams shown in Figs.~\ref{2d-pd}(a) and \ref{2d-pd}(b) in the $\alpha$-$\beta$ plane for fixed $\Omega$, and in Figs.~\ref{2d-pd}(d), \ref{2d-pd}(e), and \ref{2d-pd}(f) in the $\alpha$-$\Omega$ plane for fixed $\beta$, illustrating the LD-HD/LD-MC-HD phase boundary. See also the 3D phase diagram shown in Fig.~\ref{3d-pd-abomega} in the $\alpha$-$\beta$-$\Omega$ space, illustrating the LD-HD/LD-MC-HD phase boundary.

As $x_{\alpha} \to 0$ and $x_{\beta} \to 1$, the maximal-current segment in the LD-MC-HD phase gradually expands to occupy the entire lattice. The transition from LD-MC-HD to the MC phase occurs at $x_{\alpha}=0$ and $x_{\beta}=1$, yielding the condition
\begin{equation}
\frac{\alpha \beta}{\alpha + \beta} = \frac{1}{2}.
\label{ldmchd-and-mc-boundary}
\end{equation}
See Figs.~\ref{2d-pd}(a-c) in the $\alpha$-$\beta$ plane for fixed $\Omega$ and Figs.~\ref{2d-pd}(d-f) in the $\alpha$-$\Omega$ plane for fixed $\beta$ for the LD-MC-HD/MC boundary. See also the 3D phase diagram illustrated in Fig.~\ref{3d-pd-abomega} in the $\alpha$-$\beta$-$\Omega$ space, displaying the LD-MC-HD/MC phase boundary.

The density varies continuously across both LD-HD $\leftrightarrow$ LD-MC-HD and LD-MC-HD $\leftrightarrow$ MC transitions, as evident in Fig.~\ref{den-curr-profile1}(a), indicating that these transitions are second order.

 \section{Summary and outlook}
 \label{summary}
 
 We have thus studied an asymmetric exclusion process on a ring with weakly nonconserving Lk dynamics coupled with a finite resources. The latter dynamically controls the effective entry and exit rates of the TASEP channel. The Lk dynamics is assumed to have equal attachment and detachment rates that scale inversely with the TASEP size $L$, which ensures that the bulk hopping dynamics of TASEP can compete with the attachment-detachment process of Lk. Thus our model is a ``finite resources'' variant of an open TASEP with Lk. On the other hand, the presence of the reservoir, assumed to be without any spatial extent or internal dynamics, naturally breaks the translation invariance along the ring. Thus in that respect our model provides a generalization of a TASEP on ring with a point defect studied in Ref.~\cite{tirtha-lk1}. Thus our model connects two distinct studies in two different ways. 
 
 We have studied our model by means of MFT, supplemented by extensive MCS studies. We have obtained the stationary density profiles and the phase diagrams parametrized by the control parameters, which in this case are the parameters that define the entry and exit rate functions of the TASEP and the attachment-detachment rates (assumed equal) of the Lk. The phase diagrams reveal that our model admits {\em less} phases than an open TASEP with Lk, but more phases than a TASEP on a ring with a point defect. These differences are attributed to the interplay between finite resources and weakly nonconserving Lk dynamics in the bulk of the TASEP.  The MFT and MCS results agree very well in our model. This is in contrast to the clearly noticeable differences between the two for small $\Omega$ limit (for which the nonconserving effects due to Lk is minimal) of Ref.~\cite{tirtha-lk1}.  This can be explained by the extending the arguments developed in Ref.~\cite{tirtha-lk1}. Indeed in our case the presence of the attached particle reservoir can decorrelate the fluctuations in the TASEP lane, making the TASEP lane similar to an open TASEP, which explains good agreement between the MFT and MCS predictions. In contrast, for a TASEP on a ring with Lk in the small $\Omega$ limit, the TASEP lane by itself is effectively number conserving making the density correlation effects, neglected in the MFT,    
substantial, which causes a significant discrepancy between the MFT and MCS results.

Our studies here can be extended in a variety of ways. Instead of equal attachment and detachment rates, one can consider unequal attachment and detachment rates as used in Refs.~\cite{ef-lktasep-pre,tirtha-lk2}. It will be interesting to consider variants of the function $f$, which have distinct properties, e.g., functional forms that allow for unlimited reservoir occupation. One may further consider a point defect in the TASEP lane, which can compete with the reservoir and should have connections with the study reported in Ref.~\cite{pal-point-defect}. We hope our studies will provide further impetus along these lines. 
\appendix

 \section{EQUATIONS OF MOTION AND PARTICLE-HOLE SYMMETRY}
 \label{ph-symmetry}

 In this section, we derive the equations of motion in our system and the underlying particle-hole symmetry. We start from the discrete version of the model and switch to the continuum limit. At any bulk site $1 < j < L$ of the lattice, particle occupation number $n_{j}(t)$ at time $t$ evolves according to the following equation:
\begin{align}
\label{bulk eom}
\begin{split}
\frac{dn_{j}(t)}{dt}=&n_{j-1}(t)(1-n_{j}(t))-n_{j}(t)(1-n_{j+1}(t))\\
&+\omega(1-n_{j}(t))-\omega n_{j}(t),
\end{split}
\end{align}
where $\omega$ denotes the particle attachment and detachment rate at any bulk site, assumed to be equal. Similarly, for the boundary sites, $j=1$ and $j=L$, time-evolution equations of occupation numbers, $n_{1}(t)$ and $n_{L}(t)$, read:
\begin{eqnarray}
  &&\frac{dn_{1}(t)}{dt}=\alpha_\text{eff}(1-n_{1}(t))-n_{1}(t)(1-n_{2}(t)), \label{site1 eom} \\
  &&\frac{dn_{L}(t)}{dt}=n_{L-1}(t)(1-n_{L}(t))-\beta_\text{eff}n_{L}(t), \label{siteL eom}
 \end{eqnarray}
where $\alpha_\text{eff}$ and $\beta_\text{eff}$ are given in equations~(\ref{effective-entry-and-exit-rates}) and (\ref{f}). The system exhibits particle-hole symmetry: a particle's rightward jump corresponds to a leftward vacancy move. Similarly, particle dynamics at the boundaries and bulk (entry, exit, attachment, detachment) mirror vacancy movements in reverse. One can easily verify that equations~(\ref{bulk eom})-(\ref{siteL eom}) remain invariant under the following transformations:
\begin{eqnarray}
  &&j \leftrightarrow L-j+1, \label{tr1} \\
  &&t \leftrightarrow t, \label{tr2} \\
  &&n_{j} \leftrightarrow 1-n_{L-j+1}, \label{tr3} \\
  &&\alpha_\text{eff} \leftrightarrow \beta_\text{eff}, \label{tr4}
 \end{eqnarray}
We now assume unit geometric length of the lattice and introduce a quasicontinuous variable $x=j\epsilon \in [0,1]$, where $\epsilon=1/L$ is the lattice constant. Finally, taking the time-average of equation~(\ref{bulk eom}) together with the mean-field approximation, $\langle n_{i}(t)n_{j}(t) \rangle \approx \langle n_{i}(t) \rangle \langle n_{j}(t) \rangle = \rho_{i}(t)\rho_{j}(t)$, and replacing the discrete density $\rho_{j}(t)$ with continuous density $\rho(x,t)$, one gets the following equation ~\cite{ef-lktasep-prl,ef-lktasep-pre}:
\begin{equation}
 \label{cont bulk eq}
 \partial_{t}\rho(x,t)=-(1-2\rho)\partial_{x}\rho+\Omega(1-2\rho),
\end{equation}
where $\Omega=\omega/L$ is the total attachment or detachment rate. In the steady-state when $\partial_{t}\rho(x,t)=0$, equation~(\ref{cont bulk eq}) thus becomes equation~(\ref{eom-bulk-cont}).

 \end{document}